\documentclass[twocolumn,aps]{revtex4}

\usepackage{graphicx}% Include figure files
\usepackage{dcolumn}% Align table columns on decimal point
\usepackage{bm}% bold math
\usepackage[cmex10]{amsmath}
\usepackage{epstopdf}
\usepackage{color}

\begin{document}

\preprint{APS/123-QED}

\title{Backhopping effect in magnetic tunnel junctions: comparison between theory and experiment}

\author{Witold Skowro\'{n}ski}
 \email{skowron@agh.edu.pl}
\affiliation{Department of Electronics, AGH University of Science and Technology, Al. Mickiewicza 30, 30-059 Krak\'{o}w, Poland}

\author{Piotr Ogrodnik}
 \email{piotrogr@if.pw.edu.pl}
\affiliation{Faculty of Physics, Warsaw University of Technology, ul. Koszykowa 75, 00-662 Warsaw, Poland}
\affiliation{Institute of Molecular Physics, Polish Academy of Sciences, Ul. Smoluchowskiego 17, 60-179 Pozna\'{n}, Poland}

\author{Jerzy Wrona}
\author{Tomasz Stobiecki}
\affiliation{Department of Electronics, AGH University of Science and Technology, Al. Mickiewicza 30, 30-059 Krak\'{o}w, Poland}

\author{Renata \'{S}wirkowicz}
\affiliation{Faculty of Physics, Warsaw University of Technology, ul. Koszykowa 75, 00-662 Warsaw, Poland}

\author{J\'{o}zef Barna\'{s}}
\affiliation{Faculty of Physics, Adam Mickiewicz University, Ul. Umultowska 85, 61-614 Pozna\'{n}, Poland}
\affiliation{Institute of Molecular Physics, Polish Academy of Sciences, Ul. Smoluchowskiego 17, 60-179 Pozna\'{n}, Poland}

\author{G\"{u}nter Reiss}
\affiliation{Thin Films and Physics of Nanostructures, Bielefeld University, 33615 Bielefeld, Germany}

\author{Sebastiaan van Dijken}
\affiliation{NanoSpin, Department of Applied Physics, Aalto University School of Science, P.O.Box 15100, FI-00076 Aalto, Finland}

\date{\today}% It is always \today, today,

\begin {abstract}

We report on the magnetic switching and backhopping effects due to spin-transfer-torque  in magnetic tunnel junctions. Experimental data on the current-induced switching in junctions with MgO tunnel barrier reveal a random back-and-forth switching between the magnetization states, which appears when the current direction favors the parallel magnetic configuration. The effect depends on the barrier thickness $t_b$, and is not observed in tunnel junctions with very thin MgO tunnel barriers, $t_b$ $<$ 0.95 nm. Switching dependence on the bias voltage and barrier thickness is explained in terms of  the  macrospin model, with the magnetization dynamics described by the modified Landau-Lifshitz-Gilbert equation. Numerical simulations indicate that the competition between in-plane and out-of-plane torque components can result at high bias voltages in a non-deterministic switching behavior, in agreement with experimental observations. When the barrier thickness is reduced, the overall coupling between the magnetic layers across the barrier becomes ferromagnetic, which suppresses the backhopping effect.

\end{abstract}

%\pacs{75.47.-m, 72.25.-b}% PACS, the Physics and Astronomy
\maketitle

\section{Introduction}

Magnetic tunnel junctions (MTJs) consisting of two thin metallic ferromagnetic layers separated by an ultrathin layer of insulating material exhibit  a tunneling magnetoresistance (TMR) effect associated with the change of magnetic configuration from parallel to antiparallel alignment \cite{julliere_tunneling_1975}. The magnitude of the effect significantly depends on the insulating barrier. A very large TMR ratio has been found in MTJs with epitaxial MgO barriers \cite{yuasa_giant_2004}. In the later case, the TMR effect cannot be accounted for by the simple Julliere model, and the large TMR results rather from specific spin-filtering properties of the epitaxial MgO barrier \cite{butler_spin-dependent_2001}. Owing to large TMR ratio, the MTJs with MgO tunnel barriers are considered as highly promising systems for various applications in spintronics devices and also in information technology \cite{yuasa_giant_2007}. Indeed, MgO-based MTJs exhibiting large TMR ratio are already used as bit-cells in magnetic nonvolatile memories \cite{Kawahara_spin-transfer_2012}.

The magnetic configuration of MTJs, and thus also the corresponding resistance, can be controlled either by magnetic fields or by spin polarized currents {\it via} the spin-transfer-torque (STT) effect. Due to the STT, the MTJ-based memory cells can be switched between the two bistable states with parallel (P) and antiparallel (AP) magnetizations -- depending on the orientation of the tunneling current. It has been  shown, however,  that some random switching between the AP and P states can occur at some specific voltage conditions \cite{sun_high-bias_2009, oh_bias-voltage_2009}. This effect is now known as the backhopping phenomenon. On one side, the backhopping effect can deteriorate the memory performance \cite{min_back-hopping_2009, min_study_2010}, but on the other side,  this effect resembles the behavior of spiking neurons and thus could be used to emulate neuronal networks \cite{thomas_memristor-based_2013}.

In order to understand the backhopping effect, one should first understand its physical origin. For this purpose, we have carried out a detailed experimental study of this phenomenon in CoFeB/MgO/CoFeB MTJs with different tunnel barrier thicknesses. Apart from this, we have used the macrospin model and Landau-Lifshitz-Gilbert (LLG) equation (with the STT included \cite{slonczewski_current-driven_1996, berger_emission_1996}) to calculate MTJ stability diagrams. A similar approach had been applied successfully to metallic spin valve structures \cite{legall_state_2012, wang_bit_2012, zhou_micromagnetic_2011}. Using the STT components determined experimentally  from  the spin-torque diode measurements \cite{skowronski_influence_2013}, we are able to identify the necessary conditions for backhopping to occur. The stability analysis is supported by numerical simulations of the effect. The theoretical predictions are consistent with the experimental data on current-induced magnetization switching (CIMS) effect in our MTJs.

In section \ref{sec:experimental} we present experimental data on the current-induced magnetic switching in three samples of different MgO barriers. These data clearly reveal backhopping in two samples, while no backhopping was observed in junction with the thinnest MgO barrier. Theoretical modeling of the switching and backhopping phenomena is presented in section \ref{sec:theory}. Summary and conclusions are in section \ref{sec:summary}.

\section{Experiment}\label{sec:experimental}

To investigate the backhopping phenomenon, MTJs in the form of the following stack sequence were used (thickness in nanometers): buffer / PtMn (16) / Co$_{70}$Fe$_{30}$(2) / Ru(0.9) / Co$_{40}$Fe$_{40}$B$_{20}$(2.3) / wedge MgO(0.7 - 1.1) / Co$_{40}$Fe$_{40}$B$_{20}$(2.3) / capping. The tunnel barrier thickness varied from 0.76 to 1.01 nm. The multilayer structures were deposited in a Singulus Timaris system. After deposition, the films were annealed at 340$^\circ$C in a magnetic field of 1 T to set the exchange bias direction. Afterwards, MTJs with three different MgO barrier thicknesses were patterned using electron beam lithography, ion-milling and lift-off processes. The nanopillars had an elliptical cross-section with the short and long axis equal to  150 nm and 250 nm, respectively. The corresponding TMR ratio  varied from 110\% for 0.76 nm thick MgO tunnel barrier up to 170\% for 1.01 nm thick MgO barrier.

\begin{figure}%[!t]
	\includegraphics[width=8cm]{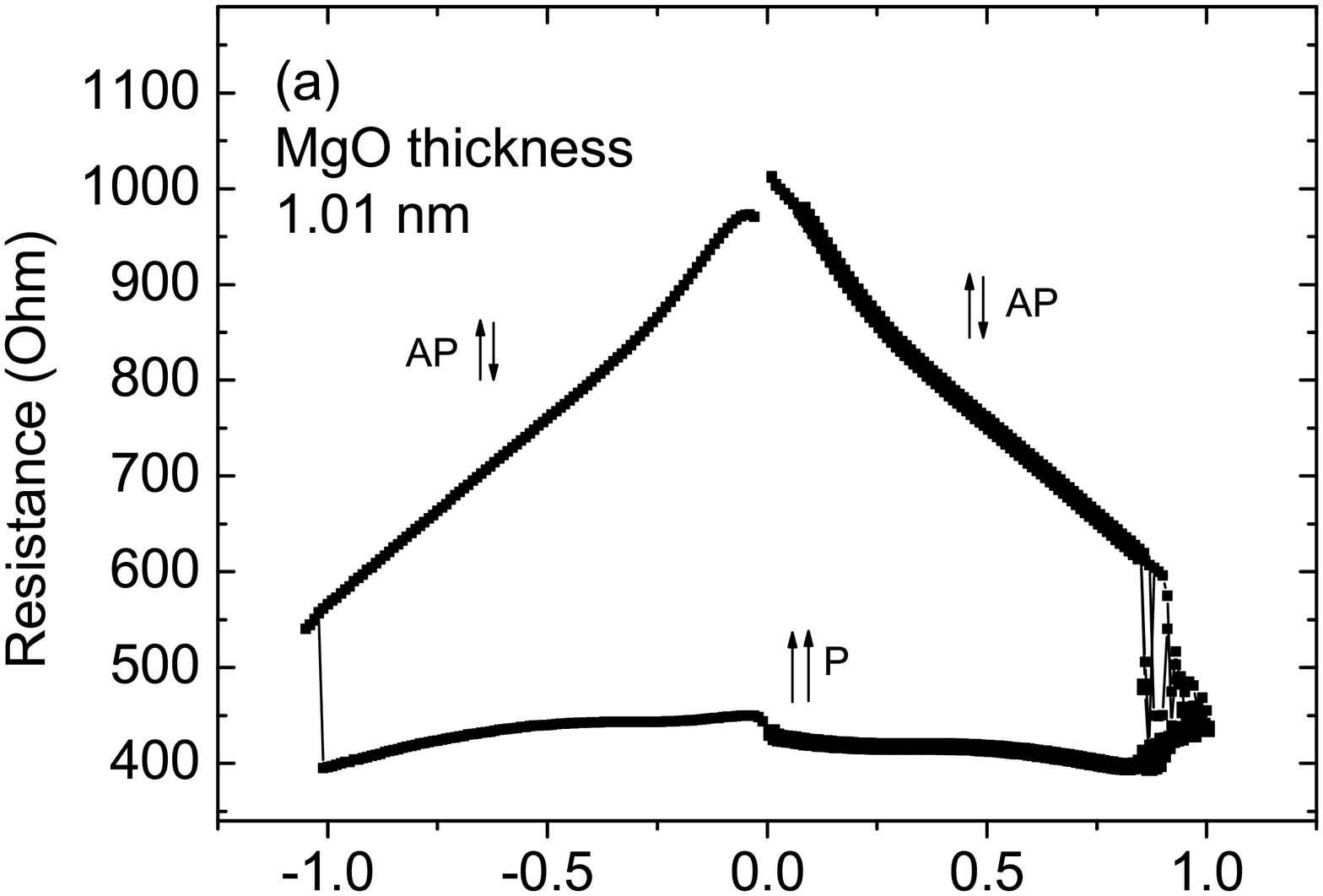}
	\includegraphics[width=8cm]{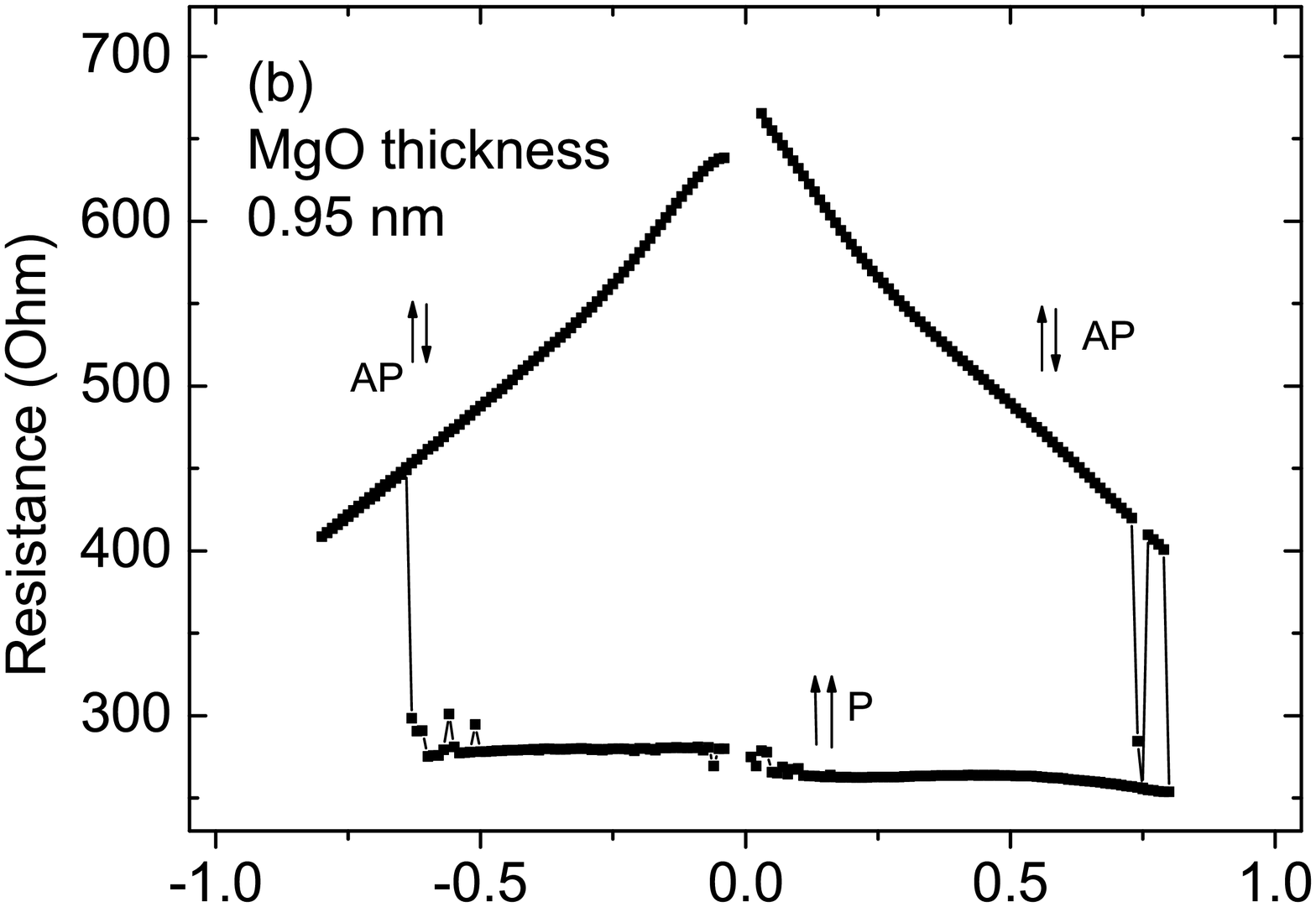}
 	\includegraphics[width=8cm]{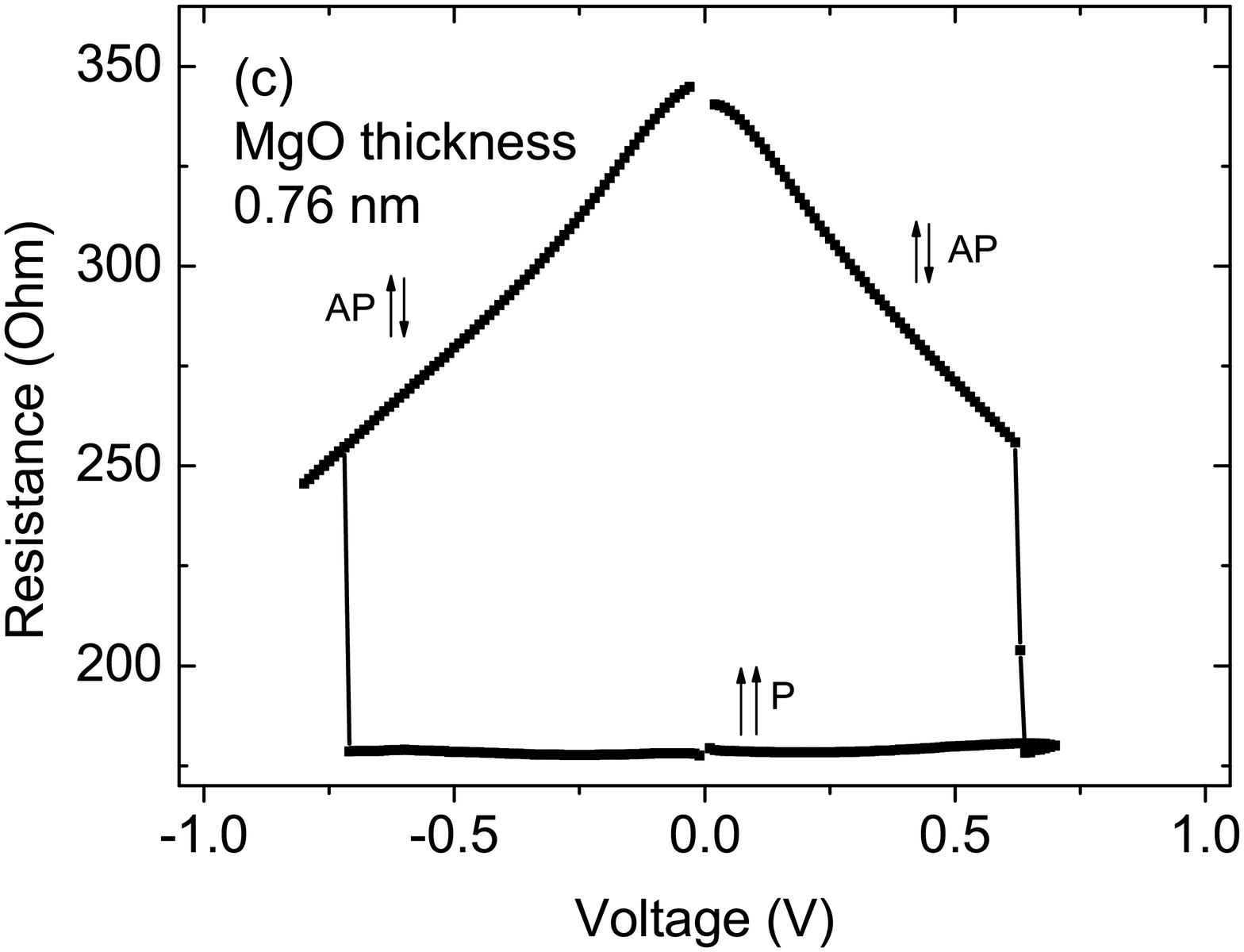}
	\caption{Experimental CIMS loops of the MTJ with 1.01 nm (a), 0.95 nm (b) and 0.76 nm (c) thick MgO tunnel barriers, measured during the voltage pulse.}% (a) and after the pulse (b) at low bias of $V_b$ = 10 mV.}
\label{fig:cims_1}
\end{figure}

The backhopping effect was observed during the CIMS measurements. In these experiments voltage pulses with a duration of 10 ms and an amplitude ranging from 0 up to $\pm$ 1 V were applied to the MTJs. The resistance of the MTJs was measured both during the switching pulse (Fig. \ref{fig:cims_1}) and after the pulse (not shown), to make sure that the final state is stable. In our case,
positive voltage indicates electron tunneling from the bottom reference layer to the top free layer.
Thus, positive voltage polarity favors switching from the AP to P magnetic state of our MTJs \cite{skowronski_interlayer_2010}, while negative voltage switches the system back from the P to AP states.

Figure~\ref{fig:cims_1}(a) presents experimental CIMS loops for the sample with 1.01 nm thick MgO tunnel barrier. For this MTJ, the application of sufficiently large positive voltage pulses, $V > 0.85$ V, induces random transitions between the AP and P states. Both these states are stable, which is confirmed by a constant junction resistance after the pulse duration. Contrary, for negative voltages only single switching events were recorded.

Figures \ref{fig:cims_1}(b) and \ref{fig:cims_1}(c) present similar CIMS loops for the MTJs with 0.95 nm and 0.76 nm thick MgO tunnel barriers, respectively. In the former case only a single backhopping event was observed for $V>0.75$ V, while in the latter case an abrupt transition from the AP to P state for positive voltage pulses was recorded, without any backhopping. As in the case shown in Fig.\ref{fig:cims_1}(a), only a single transition from the P to AP configuration was observed in both samples for negative voltage pulses.

To explain qualitatively the aforementioned behavior, we note first that the thicker the MgO tunnel barrier, the larger voltage pulse  is required for the CIMS to occur.
Magnitudes of the in-plane and out-of-plane  torques for thick barriers  are comparable. In addition, for positive bias voltage both torque components have opposite signs and therefore the competition between both torques triggers backhopping events. More specifically, as the in-plane torque  tends to switch the system to the P state, the out-of-plane torque tends to restore the AP configuration. Moreover, coupling between the free and reference layers is antiferromagnetic for thick tunnel barriers, which additionally enhances the out-of-plane torque (see Eq. \ref{eq:torques}). The sign of this coupling turns out to be important for the occurrence of backhopping. %, originating from the stray fields interactions.
For thin tunnel barriers, on the other hand, switching of the free layer occurs at smaller voltages, for which the in-plane torque dominates. Moreover,  coupling between the magnetic layers  is then ferromagnetic \cite{skowronski_interlayer_2010}, which effectively diminishes the role of the out-of-plane torque. %, originating from the interlayer exchange interactions \cite{skowronski_interlayer_2010}.

\section{Theoretical description}\label{sec:theory}

In this section  we consider theoretically the  backhopping effect observed in the MTJs, and compare theoretical predictions with the experimental observations. To do this we describe the system in terms of the macrospin model and apply the LLG equation to describe its magnetization dynamics. The LLG equation includes the effects of all effective magnetic fields, as well as of the current-induced STT.  Let us denote a unit vector along the spin moment of the free layer by  $\vec{s}$. The LLG equation for the vector $\vec{s}$  takes then the  following form:
\begin{equation}
\frac{d\vec{s}}{dt}+\alpha \vec{s} \times \frac{d\vec{s}}{dt} = \vec{\Gamma}_U + \vec{\Gamma}_P + \vec{\tau}
\label{eqn:spin_vec},
\end{equation}
where $\alpha$ is the Gilbert damping parameter, while the
right-hand side represents the total torque exerted on the magnetization vector. This torque includes contributions due to  uniaxial ($\Gamma_U$) and planar ($\Gamma_P$) anisotropies, as well as STT ($\vec{\tau}$) due to spin current flowing through the junction. In the MTJ structures, both components of STT, i.e., the in-plane and out-of-plane ones are of a similar magnitude as predicted by theory \cite{heiliger_ab_2008} and also measured experimentally \cite{sankey_measurement_2007, kubota_quantitative_2007}. Therefore, both components are important and both should be taken into account in the LLG equation on equal footing.

The total STT exerted on the magnetic moment of the free layer can be thus expressed as
\begin{equation}
\vec{\tau} = \vec{\tau}_\parallel + \vec{\tau}_\perp = \vec{\tau}_\parallel+( \vec{\tau}_{C\perp} + \vec{\tau}_{0\perp} ),
\label{eq:torques}
\end{equation}
where $\vec{\tau}_\parallel$ is the in-plane torque component, while $\vec{\tau}_{\perp}$ is the total out-of-plane torque component. The latter torque consists of pure current-induced torque $\vec{\tau}_{C\perp}$ and effective torque due to coupling between the free and reference layers, $\vec{\tau}_{0\perp}$ \cite{oh_bias-voltage_2009}. The term $\vec{\tau}_{0\perp}$ originates  from the magnetostatic field (shape dependent) as well as from the exchange interaction between the magnetic free and reference layers \cite{yang_effect_2010}.

The LLG equation can be rewritten in spherical coordinates in the dimensionless form \cite{ogrodnik_spin_2010}:
\begin{eqnarray}
\frac{d\theta}{d\tilde{\tau}} = -  h_P \cos\phi \sin\theta ( \alpha \cos\theta \cos\phi + \sin\phi ) \notag\\
 -\alpha \sin\theta \cos\theta - (h_\parallel + \alpha h_\perp )\sin\theta \notag \\
\frac{d\phi}{d\tilde{\tau}} =  h_P \cos\phi ( \alpha \sin\phi - \cos\theta \cos\phi ) \notag\\
-\cos\theta + \alpha h_\parallel - h_\perp,
\label{eqn:llg_sphere}
\end{eqnarray}
where $\tilde{\tau}$ is the dimensionless time, $h_P$, $h_\parallel$ and $h_\perp$ denote the dimensionless planar anisotropy, in-plane and total out-of-plane
torques, respectively, whereas $\phi$ and $\theta$ are the azimuthal and polar angles. The planar anisotropy constant $K_P$ is normalized here to the the uniaxial anisotropy constant $K$, $h_P=K_P/K$, while the components of STT are normalized to the product of uniaxial anisotropy
constant $K$ and thickness $t_f$ of the free layer, $h_{\parallel,\perp}=\tau_{\parallel,\perp}/2K t_f$.

In the following  the LLG equation will be used to study the phenomenon of backhopping. First, we will analyze the stability diagrams of the MTJs. Then, we will present results of full scale  numerical simulations of the CIMS loops and backhopping effect, based on the LLG equation.

\subsection{Stability analysis}

The LLG equation describes a nonlinear dynamical system and has the following general form: $d\vec{x} / dt  = f(\vec{x})$, where $f$ is a nonlinear function. To identify behavior of such a system, we linearize $f$  near the equilibrium points,  defined by the condition $f(\vec{x}_0 )=0$. Equations \ref{eqn:llg_sphere} have two main equilibrium states corresponding to the spherical angles $\theta=0$ (P configuration) and $\theta=\pi$ (AP  configuration). Thus, we linearize Eq. \ref{eqn:llg_sphere} around the AP and P configurations. This procedure is the first step of the linear stability analysis \cite{Perko_differential_1991}, which relies on investigating local properties of the dynamical system, particularly properties of equilibrium points which  may be attracting (stable) points or repulsing (unstable) points. The type of an equilibrium point is determined by the eigenvalues of the dynamical matrix of the linearized equation.

The eigenvalues of the linearized form of Eq. \ref{eqn:llg_sphere} have been calculated earlier for metallic spin valves, but without considering  the bias-dependent out-of-plane torque \cite{bazaliy_current-induced_2004}. These results can be easily adopted to the situation studied in this paper. Accordingly, the four eigenvalues (one pair for each equilibrium state) have the following form:
\begin{widetext}
 \begin{equation}
 \mu^{P}_{1,2} = -\frac{1}{2}\left[ 2h_\parallel+\alpha(2+h_P+2h_\perp) \pm \sqrt{4[\alpha h_\parallel (2+h_P+2h_\perp)-\alpha^2 h_\parallel^{2}-1-h_P-(2+h_P)h_\perp-h^2_\perp]+\alpha^2h^2_P} \right]
\end{equation}
\begin{equation}
\mu^{AP} _{1,2}= -\frac{1}{2} \left[ 2\alpha-2h_\parallel+\alpha h_P - 2\alpha h_\perp \pm \sqrt{4\left[      h_P h_\perp - 1 - \alpha^2 h^2_\parallel - h_P - \alpha h_\parallel h_P -h^2_\perp \right] +8\left[h_\perp+ \alpha h_\parallel h_\perp - \alpha h_\parallel \right] + \alpha^2 h^2_P }\right].
\end{equation}
\end{widetext}

\begin{figure}[!t]
      \includegraphics[width=\columnwidth]{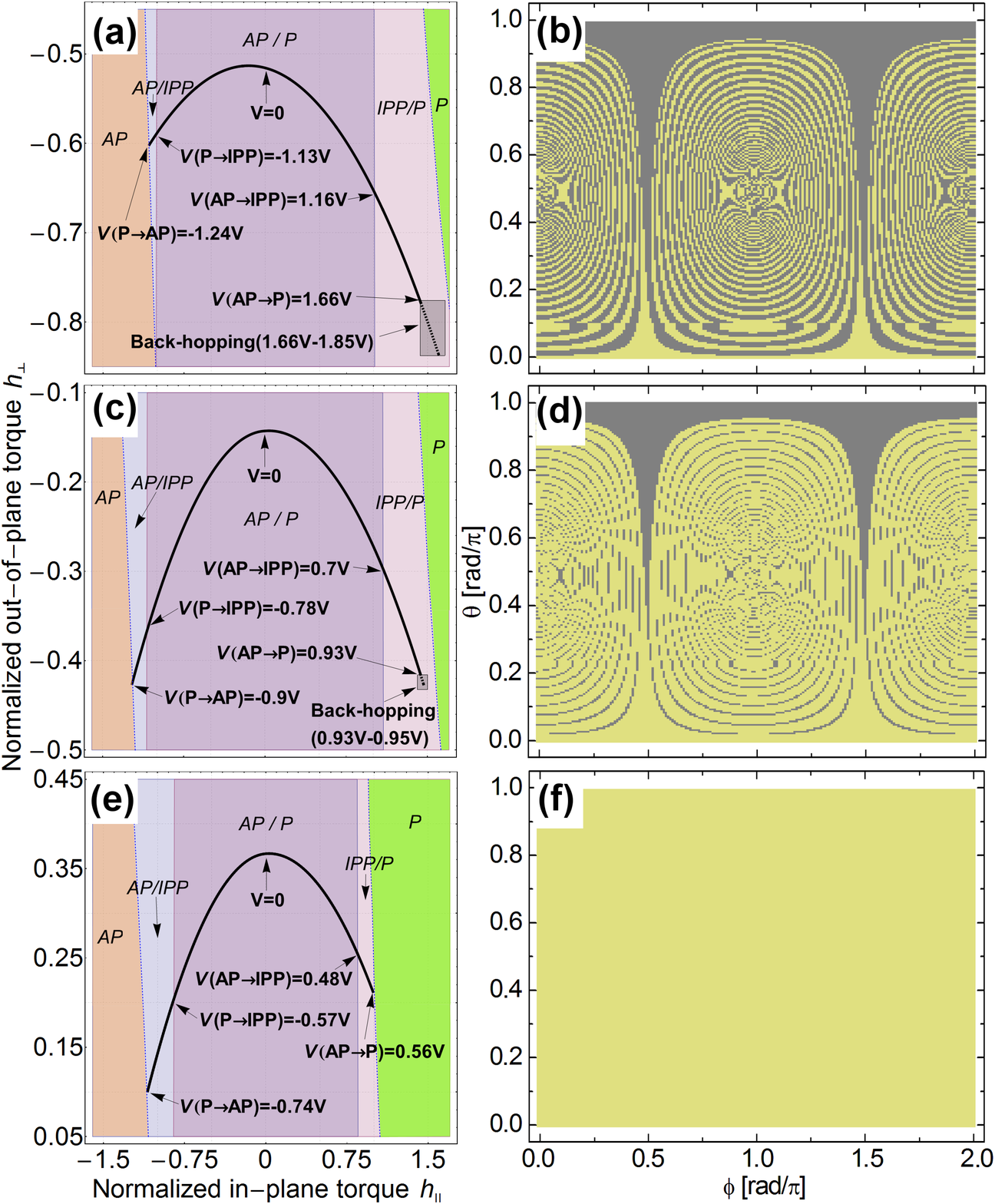}
	\caption{(Color online) Stability diagrams corresponding to the three experimental samples with different MgO thickness: 1.01 nm (a), 0.95 nm (c), 0.76 nm (e). The corresponding parameters are: magnetization saturation ($M_S=0.9$T, $1.0$T, $1.0$T), room-temperature anisotropy field ($H_K = 64$Oe, $70$Oe, $120$Oe), damping coefficient ($\alpha = 0.014, 0.015, 0.020$), and interlayer coupling field ($H_C = -33$Oe, $-10$Oe, $+44$Oe), respectively. Different areas correspond to the different stable/bistable solutions of the LLG equation:  P -- parallel state,	AP -- antiparallel state, IPP -- in-plane precessions (close to the AP state in the right areas, positive voltage, and close to the P state in left areas, negative voltage). The solid line corresponds to STT taken from experiment.
Right panel presents basins of attractions for the samples with MgO thickness 1.01 nm (b), 0.95 nm (d) and 0.76 nm (f), calculated for voltages 1.7 V, 0.95 V and 0.6 V, respectively.  Yellow (bright in print) and grey (dark in print) colors correspond to initial conditions resulting in the stable P and IPP states, respectively.} \label{fig:bassin1}
\label{fig:diag1}
\end{figure}

The sign and character (real or imaginary) of the above eigenvalues determine the stable or unstable character of the P and AP states. This allows us to create the stability diagrams for both magnetization configurations, shown in Figs \ref{fig:diag1}(a,c,e) for the three investigated junctions corresponding to different MgO barrier thicknesses. The diagrams are plotted in the space of  normalized in-plane and out-of-plane torque components. Different areas in these diagrams correspond to different types of stable solutions of the LLG equation. In general, one can distinguish five cases: (i) the P and AP states are both stable, (ii) only the P state is stable, (iii) only the AP state is stable, (iv) the P state is stable while stationary in-plane precessions occur around the unstable AP configuration, and (v) the AP state is stable while stationary in-plane precessions occur around the unstable P configuration. The stable in-plane oscillations, marked in Figs \ref{fig:diag1}(a,c,e) as IPP states, arise from loss of stability of the P or AP state. When a stable point changes to an unstable repulsing point, then the stable precessional solution may appear. The mechanism is known as a supercritical Hopf bifurcation \cite{Perko_differential_1991}, and it has been checked that it is present also in the considered system. Contrary, the IPP state loses its stable character due to homoclinic bifurcation \cite{Perko_differential_1991}. In Figs \ref{fig:diag1}(a,c,e), the stability of the IPP state has been determined by the numerical calculation.

The black solid lines in Figs \ref{fig:diag1}(a,c,e) correspond to the STT components ($h_\parallel$ and $h_\perp$) taken from experimental data at different bias voltages \cite{skowronski_influence_2013}. Since we were unable to measure the STT components for voltages higher than $0.4$ V, we assumed that the in-plane ($\tau_\parallel$) and out-of-plane ($\tau_{C\perp}$) torques depend linearly and quadrically on the bias voltage \cite{jia_nonlinear_2011}: $\tau_\parallel(V)=aV$ and $\tau_{C\perp}(V)=bV^2+cV$. Therefore, the out-of-plane torque can be expressed through the in-plane component as: $\tau_\perp(\tau_\parallel)=b {\tau_\parallel}(V)^2/a^2 + c \tau_\parallel (V)/a+\tau_{0\perp}(V=0)$, where $a,b,c$ are parameters taken from fitting to experimental results. %The solid black line in Figs {\ref{fig:diag1}}(a,c,e) is described by the same equation with normalized $h_{\parallel (\perp)}$ instead of $\tau_{\parallel (\perp)}$.

In each of the bistable regions in Figs \ref{fig:diag1}(a,c,e), two types of stable solutions of Eq. \ref{eqn:llg_sphere} are possible: P or AP, P or IPP, and AP or IPP. Note, the IPP states associated with P and AP configurations are different, as already mentioned above.  The influence of interlayer coupling (contributing to $h_{0\perp}$) is clearly visible when comparing Figs \ref{fig:diag1}(a,c,e). In Fig.\ref{fig:diag1}(a) the overall coupling between the free and reference layers is antiferromagnetic, so the out-of-plane torque (see the  black solid line) for $h_\parallel = 0$  is negative. Both the antiferromagnetic coupling and out-of-plane torque favor then the AP state. This is the reason why a larger value of in-plane torque (achievable at higher voltage) is needed for the AP to P switching in this sample. The coupling in Fig.\ref{fig:diag1}(c) is still antiferromagnetic but its magnitude is reduced. Contrary,  in Fig. \ref{fig:diag1}(e) the net interlayer coupling has ferromagnetic character, which results in positive out-of-plane torque for $h_\parallel=0$.
The above discussion also accounts for one feature of the diagrams in Figs \ref{fig:diag1}(a,c,e), namely the fact that the part of solid line (experimental torque) laying in the IPP/P bistable region covers the widest voltage range (about 0.5 V) for the MTJ with the thickest MgO tunnel barrier, $t_b$ = 1.01nm (see Fig. \ref{fig:diag1}(a)), whereas very narrow bistable range is simulated for the thinnest MgO tunnel barrier, $t_b$ = 0.76nm.

Magnitudes of the competing in-plane and out-of-plane torque components have a significant influence on the probability of backhopping. This backhopping appears for voltages above the voltage at which switching from the AP (or more correctly from IPP) to P states appears (see Figs \ref{fig:diag1}(a,c)). However, experimental data show that the multiple backhopping appears only for the MTJ with the thickest MgO tunnel barrier, when the out-of-plane torque is sufficiently strong. We note that  the backhopping appears between the IPP (close to AP configuration) and P states. In the MTJ with the thinner barrier, $t_b=0.95$ nm, only a single backhopping event was observed experimentally, while no  backhopping was noticed for the thinnest barrier, $t_b=0.76$ nm. This  is because  the in-plane torque in MTJs with thinner MgO barriers is larger than the out-of-plane one, and thus the switching from the AP to P states occurs at a relatively small bias voltage, for which the out-of-plane torque is not able to switch the system back to the IPP state. The IPP state occurs rather near the unstable AP configuration and when the MTJ is switched to the P state, the IPP state is not achievable anymore (or very rarely). The bistable regions in the  stability diagrams indicate that transition between the corresponding two solutions are allowed, but say nothing about their probabilities. To study this problem in more details, we have calculated the corresponding basins of attraction, shown in Figs \ref{fig:diag1}(b,d,f).

Numerical calculations of the attraction basins have been performed for  V=0.01 V (low voltage state, not shown) as well as at voltages slightly greater than AP$\rightarrow$P switching voltages: $V=1.7$ V (for 1.01 nm MgO), $V=0.95$ V (0.95 nm MgO), and $V=0.6$ V (0.76 nm MgO). The results, shown in Figs \ref{fig:diag1}(b,d,f), are consistent with the stability diagrams from Figs \ref{fig:diag1}(a,c,e). In Figs \ref{fig:diag1}(b,d) there are two possible solutions: P corresponding to the yellow areas (bright in print) and IPP corresponding to the gray areas (dark in print). For the junction with thick MgO barrier, Fig.\ref{fig:diag1}(b), there is a large probability of IPP solution even after AP $\rightarrow$P switching. There are wide grey stripes near the P state ($\theta=0$), which correspond to the IPP solution.
Contrary, Fig. \ref{fig:diag1}(d) is dominated by yellow color, while the gray stripes are very narrow. This indicates that after switching to the P state, transition back to the IPP state is much less probable, though still possible with some small probability, which is in agreement with our experimental observations as well as with numerical simulations to be described later on.
This difference in attraction basins holds at low voltage state (V=0.01V) as well (not shown).
Thus, the P state is more stable for junctions with thinner MgO barriers. The difference is due to different magnitudes of the antiferromagnetic coupling, as already discussed above.

For the MTJ with the thinnest MgO barrier, the overall coupling between the free and reference layers is strongly ferromagnetic. This results in a very narrow positive voltage range for which IPP oscillations can occur, in contrast to negative bias voltages, where IPP oscillations are present in a wider range. However, the main difference between the MTJ with the thinnest MgO barrier and thicker barriers is that the IPP oscillations do not exist after the switching to the P configuration, thus, the backhopping is not possible.
This is clearly visible in the corresponding attraction basin (see Fig. \ref{fig:diag1}(f)), where all the area is yellow (bright in print), indicating that the P state is stable and no backhopping to the IPP state can appear.

\begin{figure}%[t]

	\includegraphics[width=8cm]{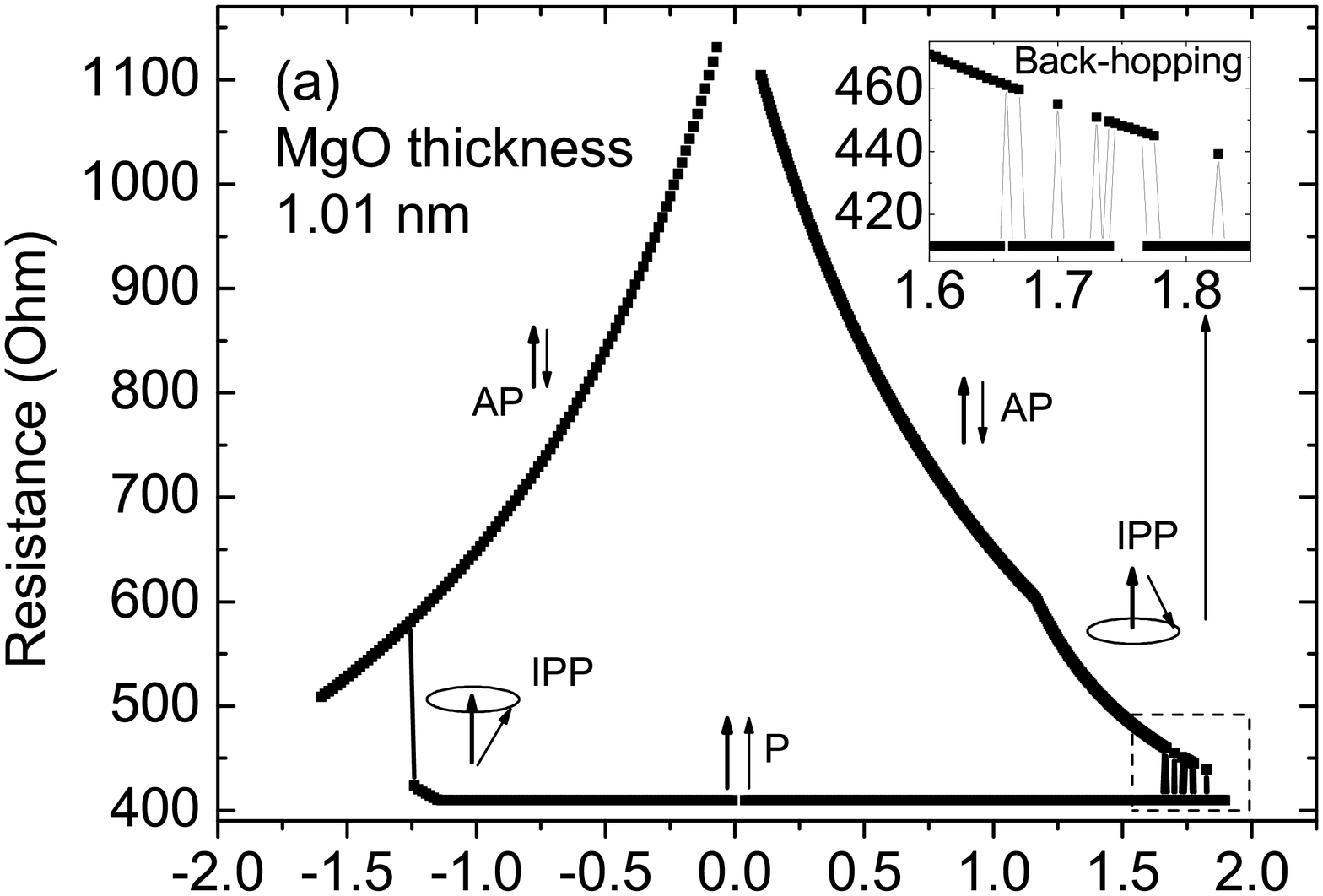}
	\includegraphics[width=8cm]{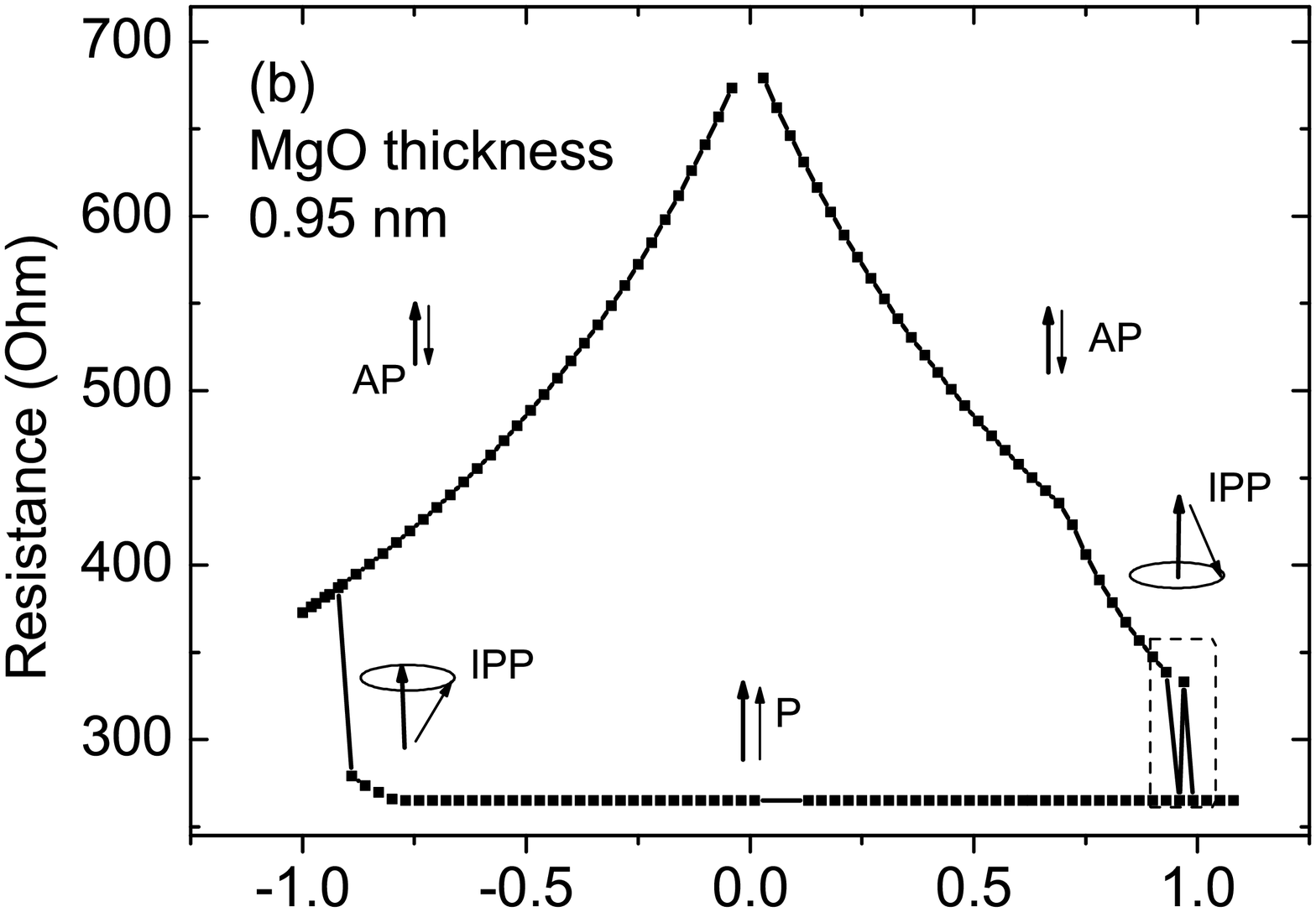}
	\includegraphics[width=8cm]{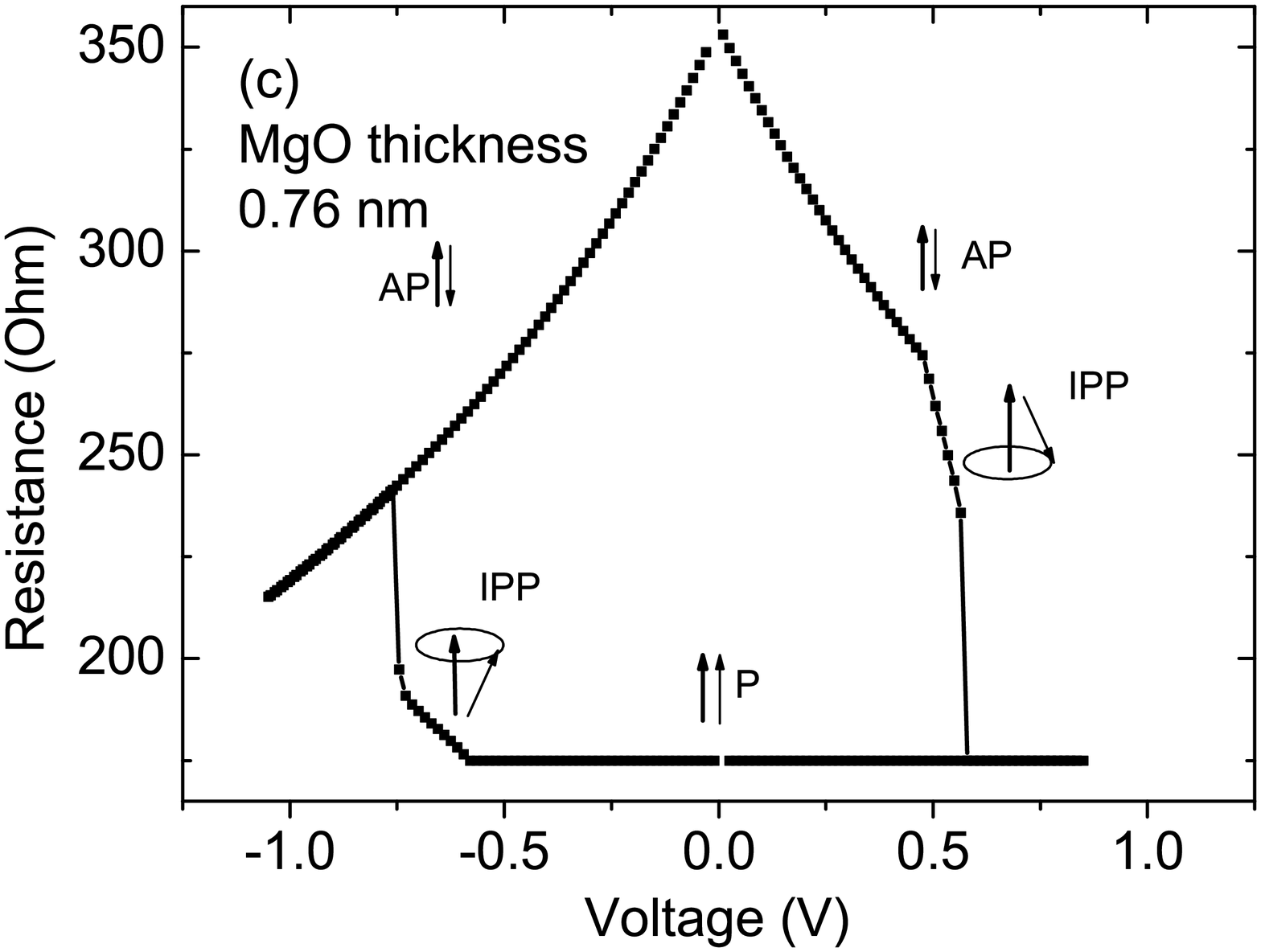}
	\caption{Numerical simulations of the CIMS loops for three MTJs  with different MgO tunnel barrier thicknesses. The inset in (a) shows the region of the backhopping instability.}
\label{fig:diag2}
\end{figure}

\subsection{Numerical simulations}

The conclusions we arrived at when analyzing the stability conditions of the P and AP  states of the MTJs, have been supported by numerical full-scale simulations. To solve the LLG equation, Eq.\ref{eqn:llg_sphere}, we used the fourth-order Runge-Kutta algorithm. From the time evolution of the polar angle $\theta$, we calculated its mean value $\theta_{m}$ and then the junction resistance according to the formula: $R(\theta_{m}) = (R_P+R_{AP})/2 + [(R_P - R_{AP})/2] \cos(\theta_{m})$, where $R_{AP}$ and $R_{P}$  are the resistances in the AP and P states, respectively. We assumed that initial conditions deviate from $\theta=0,\pi$ due to thermal fluctuations estimated for a temperature of 300K.\cite{mayergoyz2009nonlinear}

The parameters used in simulations were taken from our previous works \cite{skowronski_interlayer_2010, serrano-guisan_inductive_2011}, whereas STT components were measured using the spin-torque diode effect \cite{skowronski_influence_2013}, and are given in the caption to Fig. \ref{fig:diag1} and also in the main text. Results on the simulations of CIMS loops  are presented in Fig. \ref{fig:diag2} for all three junctions investigated experimentally. When comparing  the experimental results of Fig. \ref{fig:cims_1} with the theoretical results of Fig. \ref{fig:diag2}, one finds a good qualitative correspondence. The switching from AP to P state occurs {via} IPP states, which are close to the AP configuration, so the corresponding difference in the resistance is only weakly resolved in simulations. Moreover, for the thickest barrier there is a clearly resolved multiple backhopping effect. For the thinner barrier only a single backhopping event was detected, while no backhopping was found for the thinnest tunnel barrier, in agreement with experimental data. For negative voltages only a single switch from the P to the AP state occurs and a narrow voltage range with IPP precessions close to the P state exists.
Some  discrepancy in the experimental and theoretical switching voltages may be ascribed to additional thermal effects in the experiment, which could lower the critical current density, and which were not taken into account in the numerical simulations.
Moreover, the numerical simulations also show that the transition from  antiferromagnetic to  ferromagnetic effective  coupling between the free and  reference layers, which depends on the MgO barrier thickness, is an important source for the backhopping effect.

\section{Summary and conclusions}\label{sec:summary}

In summary, we have studied experimentally the CIMS loops in MgO-based MTJs with different MgO barrier thicknesses. The experimental data clearly showed that backhopping occurs for thicker tunnel barriers. No backhopping was found in the MTJ with very thin tunnel barrier, where the effective interaction between the free and reference layers was ferromagnetic.

Using the macrospin model and  modified Landau-Lifshitz-Gilbert equation containing the  STT terms, we analyzed the stability conditions as well as performed numerical simulations of the CIMS loops. Assuming the  experimentally determined parameters of the  MTJs, including the STT components,  magnetic anisotropy, effective damping constant, and the interlayer  coupling, we showed that backhopping between the P and AP states can be rationalized by a competition between the in-plane and out-of-plane torque components. Backhopping occurs when both torques have similar magnitude (which is the case near the switching voltage of relatively thick tunnel barriers) and opposite signs. Because antiferromagnetic coupling between the free and reference layers increases the magnitude of the out-of-plane torque, it enhances the tendency towards backhopping. For very thin barrier, on the other hand, the in-plane torque dominates near the   switching voltage and, consequently, abrupt magnetization switching without backhopping is observed.

\section*{Acknowledgement}
This project is supported by the NANOSPIN Grant No. PSPB-045/2010 from Switzerland through the Swiss Contribution and by the Polish National Science Center Grants No. NN515544538 and No. DEC-2012/04/A/ST3/00372. Numerical calculations were supported in part by PL-Grid Infrastructure. W.S. acknowledges the Foundation for Polish Science MPD Programme co-financed by the EU European Regional Development Fund. G.R. acknowledges the DFG funding (Grant No. RE 1052/22-1). S.v.D. acknowledges financial support from the Academy of Finland (Grant No. 260361).

%\bibliographystyle{nature}
%\bibliography{Skowronski_library}

\end{document}